# Integrated Digital Inverters Based on Two-dimensional Anisotropic ReS$_2$ Field-effect Transistors

Erfu Liu[1]*, Yajun Fu[1]*, Yaojia Wang[1], Yanqing Feng[1], Huimei Liu[1], Xiangang Wan[1], Wei Zhou[1], Baigeng Wang[1], Lubin Shao[1], Ching-Hwa Ho[4], Ying-Sheng Huang[5], Zhengyi Cao[6], Laiguo Wang[6], Aidong Li[6], Junwen Zeng[1], Fengqi Song[1], Xinran Wang[7], Yi Shi[7], Hongtao Yuan[2,3], Harold Y. Hwang[2,3], Yi Cui[2,3], Feng Miao[1] & Dingyu Xing[1]

**Semiconducting two-dimensional transition metal dichalcogenides are emerging as top candidates for post-silicon electronics. While most of them exhibit isotropic behavior, lowering the lattice symmetry could induce anisotropic properties, which are both scientifically interesting and potentially useful. Here, we present atomically thin rhenium disulfide (ReS$_2$) flakes with unique distorted 1T structure, which exhibit in-plane anisotropic properties. We fabricated mono- and few-layer ReS$_2$ field effect transistors, which exhibit competitive performance with large current on/off ratios (~$10^7$) and low subthreshold swings (100 mV dec$^{-1}$). The observed anisotropic ratio along two principle axes reaches 3.1, which is the highest among all known two-dimensional semiconducting materials. Furthermore, we successfully demonstrated an integrated digital inverter with good performance by utilizing two ReS$_2$ anisotropic field effect transistors, suggesting the promising implementation of large-scale two-dimensional logic circuits. Our results underscore the unique properties of two-dimensional semiconducting materials with low crystal symmetry for future electronic applications.**

[1] National Laboratory of Solid State Microstructures, School of Physics, Collaborative Innovation Center of Advanced Microstructures, Nanjing University, Nanjing 210093, China.

[2] Geballe Laboratory for Advanced Materials, Stanford University, Stanford, California 94305, USA.

[3] Stanford Institute for Materials and Energy Sciences, SLAC National Accelerator Laboratory, Menlo Park, California 94025, USA.

[4] Graduate School of Applied Science and Technology, National Taiwan University of

Science and Technology, Taipei 106, Taiwan.

[5] Department of Electronic and Computer Engineering, National Taiwan University of Science and Technology, Taipei 106, Taiwan.

[6] Department of Materials Science and Engineering, College of Engineering and Applied Sciences, Nanjing University, Nanjing 210093, China.

[7] School of Electronic Science and Engineering, Nanjing University, Nanjing 210093, China.

* These authors contributed equally to this work.

Correspondence and requests for materials should be addressed to F. M. (email: miao@nju.edu.cn), H.T.Y. (email: htyuan@stanford.edu), or B. W. (email: bgwang@nju.edu.cn).

The scaling down of conventional metal-oxide-semiconductor field-effect transistors (MOSFETs) is approaching the limit of miniaturization due to severe short-channel effects[1]. One approach to solve this problem is searching for atomically thin semiconductors, such as semiconducting two-dimensional (2D) materials[2-6]. Among the potential candidates, the family of transition metal dichalcogenides (TMDs)[7,8] ($MX_2$, where M denotes a transition metal and X denotes a chalcogen) has attracted the most attention due to both its rich physics and tremendous potential for application. They are expected to exhibit a wide range of electronic structures and exotic transport properties arising from the various electron configurations of transition metals, such as superconductivity[9-11], half-metallic magnetism[12], and charge density waves[13-15].

Lattice structure and symmetry are vital in determining materials' fundamental properties. Most studied 2D TMDs exhibit isotropic behavior due to high lattice symmetry; however, lowering the symmetry in TMDs could induce interesting anisotropic properties of both scientific and technologic importance. For example, in $WTe_2$, which is a semi-metallic TMD with a distorted lattice structure, a large non-saturating magnetoresistance effect along a principle axis (the direction of W chain formation) has been recently reported and has attracted considerable attention[16]. In contrast, semiconducting 2D TMDs with low symmetry have remained unexplored.

As the last stable element discovered in nature, the transition metal rhenium (Re) exhibits a wide range of oxidation states and crucial applications, such as superalloys and catalytic reforming. Its sulfide $ReS_2$ is a semiconducting TMD that was recently discovered to have weak interlayer coupling and a unique distorted 1T structure[17]. As shown by the side view and top view of the mono-layer $ReS_2$ crystal structure in Fig. 1a, it consists of two hexagonal planes of S atoms (yellow) and an intercalated hexagonal plane of Re atoms (blue) bound with S atoms forming chains of parallelogram-shaped Re4 clusters. As highlighted by the red arrows in the top view, there are two principle axes, the b- and a-axes, which correspond to the shortest and second-shortest axes in the basal plane. They are 61.03° or 118.97° apart, and the b-axis corresponds to the direction of the Re-Re atomic chain formed. Thus, $ReS_2$ offers an ideal material candidate for studying semiconducting 2D TMDs with low symmetry.

In this work, we systematically investigate the in-plane anisotropic properties of atomically thin $ReS_2$. In addition to highly competitive field effect transistor (FET) performance, we also observe a large anisotropic ratio along its two principle axes, which is the highest among all experimentally investigated 2D layered materials. We also successfully demonstrate a digital inverter with good performance by integrating two anisotropic ReS$_2$ FET devices. Our results suggest that the lattice orientation can be used as an alternative design variable to tune device properties and optimize circuit performance in future 2D integrated circuits based on anisotropic semiconducting materials.

Results

**Material characterizations and *ab initio* calculations.** Figs. 1b and 1c show typical optical microscopy and AFM images of a mono-layer ReS$_2$ film (see Methods for the material details), respectively. The inset of Fig. 1c shows that the thickness of the mono-layer film is approximately 0.8 nm. In this study, we focus on mono- and few-layer ReS$_2$, with thicknesses ranging from 0.8 to 5 nm (1-7 layers with interlayer spacing of approximately 0.7 nm). Micro Raman scattering experiments were performed on mono-layer, five-layer and bulk ReS$_2$. Due to the low symmetry of ReS$_2$, 18 Raman modes were observed as shown in Fig. 1d. The peak positions of our three samples are close to each other, same to the reported results[17]. In Fig. 1d we labeled two low frequency A$_g$-like modes (located at 136.8, 144.5 cm$^{-1}$) corresponding to the out-of-plane vibrations of Re atoms and four E$_g$-like modes (located at 153.6, 163.4, 218.2, 238.1 cm$^{-1}$) corresponding to the in-plane vibrations of Re atoms. The rest 12 higher frequency Raman modes are vibrations mainly from lighter S atoms[18]. The peak intensity ratio could be dependent on the polarization due to the low lattice symmetry[19]. The band structures of mono-, tri-, and five-layer ReS$_2$ calculated by an *ab initio* method are shown in Fig. 1e, indicating the nature of a direct band gap semiconductor without the indirect-to-direct band gap transition observed in other TMDs[3]. The overall band topology does not significantly change, with only a minor band gap shortening from mono-layer (1.44 eV), tri-layer (1.4 eV) to five-layer (1.35 eV).

**Field effect transistor performances.** We first examine the device performances of mono- and few-layer ReS$_2$ FET devices (see Supplementary Note 1 for details). The inset of Fig. 2a presents the optical image of a typical mono-layer device. The FET transfer curve was obtained by monitoring the source-drain current $I_{ds}$ while sweeping the back gate voltage $V_{bg}$. A fixed 100 mV source-drain bias voltage $V_{sd}$ was applied across the channel during all of the measurements. We first measured a couple of mono-layer ReS$_2$ devices, all of which behaved as excellent n-type FET devices with the back gate swept between -50 V and +50 V. As shown in Fig. 2a, the current on/off ratio reached 10$^7$, which meets the requirement of many applications, such as digital logic computation. When the back gate voltage reached +50 V, the $I_{ds}$ approached saturation. Few-layer ReS$_2$ transistor devices, ranging from bi-layer to seven-layer, were studied as well. Similar n-type FET behavior but a slightly larger on/off ratio was observed. The results from a tri-layer device with the same back gate sweep range are shown in Fig. 2a for comparison. We note that all of the above measurements were carried out under vacuum (approximately 10$^{-5}$ mbar) at room temperature, and ambipolar behavior can be observed by introducing the ionic liquid gating technique (see Supplementary Note 2 for details). Based on the transfer curves, we can also measure the subthreshold swing. For the mono-layer ReS$_2$ FET device shown in Fig. 2a, the subthreshold swing is approximately 310 mV per decade. The subthreshold swing becomes steeper for thicker flakes, reaching approximately 100 mV per decade for the tri-layer device shown in the same figure. The measured subthreshold swing values approach those measured in top-gated MoS$_2$ transistors[7] and outperform black phosphorus devices[5]. This can be further improved by increasing the gate capacitance after introducing thinner high-κ dielectric materials.

The mono- and few-layer ReS$_2$ FET devices (with Ti/Au electrodes) have also shown good contact behavior. As shown in Fig. 2b, for a typical mono-layer device, the source-drain current $I_{ds}$ varies linearly with the bias $V_{ds}$ in the ±200 mV range at different back gate voltages ($V_{bg}$ = -20 V, 0 V, 20 V, and 50 V respectively), indicating ohmic contact behavior in the n-doped regime under consideration (see Supplementary Note 3 for more details).

We further extracted the mobility of all of the FET devices we measured and studied its dependence on the number of layers of the ReS$_2$. The mobility was determined by taking the steepest slope in the two-terminal $I_{ds}$-$V_{bg}$ curves. The results for 17 devices (from mono- to seven-layer) are plotted in Fig. 2c. Similar to the reported results based on MoS$_2$ thin flakes[20,21], the device mobility generally increases with an increasing number of layers, implying that the electrons likely independently transport through different layers in thin ReS$_2$ flakes. The mobility of a mono-layer device varies between 0.1 and 2.6 cm$^2$V$^{-1}$s$^{-1}$, and the highest mobility value obtained is 15.4 cm$^2$V$^{-1}$s$^{-1}$ for a six-layer device. These values were measured without any material treatment or device optimization. With further improvement in crystal quality, post-fabrication treatment[22] and dielectric engineering[7,22], considerable enhancement of mobility to approach application requirements should be expected. Here, the scattering of mobility values for flakes with the same number of layers could be induced by device quality variations or the anisotropic property, as will be discussed below.

**Anisotropic properties of $ReS_2$.** We now focus on the in-plane anisotropic properties of mono- and few-layer $ReS_2$ flakes induced by low lattice symmetry. As described in Fig. 1a, the a- and b-axes are the two directions with the shortest axes in the basal plane, thus differentiating them from other lattice orientations. Such a feature is often associated with the in-plane anisotropic properties of structural stiffness and electronic transport. Interestingly, during our experiments, exfoliated thin $ReS_2$ flakes commonly appear in a quadrilateral shape with inner angles of approximately 60° or 120°, as shown by a typical flake in Fig. 3a. Further results of all inner angles measured on over 20 thin flakes is shown in Fig. 3b, with more than 60% of the specimens showing these two angles. The inner angles of 60° and 120° match the angles between the a- and b-axes (118.97° or 61.03°) with high accuracy. This can be readily explained by the fact that the breaking strength is the weakest along these two axes, which are the two most strongly bonded orientations[17,23], and suggests that two sides of the quadrilateral shape with 60° or 120° inner angles correspond to the a- and b-axes, respectively.

The in-plane stiffness anisotropy-induced quadrilateral shape makes it feasible to determine the two axes of $ReS_2$ flakes via convenient transport measurements instead of other highly sophisticated tools, such as scanning tunneling microscopy (STM). Early studies on bulk materials reported that the b-axis is more conductive than other crystalline orientations[24,25]. We then patterned the electrodes to be perpendicular to two sides (A and B directions) of a quadrilateral-shaped five-layer $ReS_2$ with a 60° inner angle (as shown in the top inset of Fig. 3c). The transfer curves in two directions are shown in Fig. 3c, and anisotropic FET behavior was observed. The B direction appears to be noticeably more conductive than the A direction, suggesting that the A and B directions correspond to the a- and b-axes, respectively. The conductance ratio of the two directions is gate dependent, with values of approximately 8.2 when $V_{bg}$ = -50 V and approximately 2 when $V_{bg}$ = 50 V. To evaluate the influence of contact resistance, we performed four-terminal measurements (see Supplementary Note 3 for more details). The four-terminal results of the same device are shown in the lower inset of Fig. 3c, indicating that the anisotropy arises from the intrinsic properties of ReS$_2$.

We further systematically studied the anisotropic properties of thin ReS$_2$ flakes through angle-resolved transport measurement. As shown in the inset of Fig. 3d, a six-layer device was fabricated with 12 electrodes (5 nm Ti/50 nm Au) evenly spaced at 30º apart. We measured the transfer curves of each pair of diagonally positioned electrodes separated by 4.5 μm at 180º apart and extracted the renormalized field effect mobility of each direction, with the results plotted in Fig. 3d (red dots) in polar coordinates. Measurements on each pair of electrodes lead to two data points that are 180º apart by swapping source-drain current directions. Here, we define the direction with the lowest mobility to be the 0º (or 180º) reference. The field-effect mobility is highly angle-dependent, with the largest value in the direction of 120º (or 300º), which is 60º from the direction with the lowest value. The anisotropic ratio of mobility $\mu_{max}/\mu_{min}$ is approximately 3.1, which is noticeably larger than that reported in other 2D anisotropic materials, such as 1.8 for thin-layer black phosphorus[26].

To fully understand our observations, we calculated the effective mass and mobility of mono-layer ReS$_2$ along three crystalline orientations (a-axis, b-axis and perpendicular to the a-axis) by using an *ab initio* technique (see Supplementary Note 4 for details), with the results plotted in the same graph (blue dots with right axis). Here, we set the direction with the lowest mobility (a-axis) to be the 0º (or 180º) reference as well. By comparing with experimental data, the calculation results offer a qualitative explanation that for the device with polar coordinates, the direction of 0º (or 180º) approaches the a-axis with the lowest mobility, 120º (or 300º) approaches the b-axis with the highest mobility, and 90º (or 270º) approaches the direction perpendicular to the a-axis with a moderate mobility value between the two extremes. The quantitative discrepancies between the experimental and theoretical results imply that with continuous improvement of sample quality and material engineering, it is possible to achieve intrinsic properties that approach the real potential of ReS$_2$ in high-performance device applications.

**Digital inverter based on anisotropic ReS$_2$ field effect transistors**. Anisotropic ReS$_2$ FET devices with expected high mobility could have important applications in future nanoscale electronics, especially on 2D logic circuits[27-31] with demanding requirements of scalability (< 10 nm) and large-scale integration. In contrast to conventional complementary metal-oxide-semiconductor (CMOS) technologies, tailoring material properties and integrated circuit design variables (such as the channel width/length ratio)[32] to optimize circuit performance on such a rigorous scale will remain a significant challenge. In this context, the lattice orientation could be used as an alternate design variable to tune device transport properties and optimize circuit performance in future 2D integrated circuits based on anisotropic semiconducting materials.

As a simple example, we successfully demonstrated a ReS$_2$-based prototype logic device, a 2D digital inverter. As schematically shown in Fig 4a, the inverter can be fabricated by combining two anisotropic ReS$_2$ FETs along the a- and b-axes (with a Re atomic chain highlighted in red). In our experiment, a quadrilateral-shaped few-layer ReS$_2$ flake with a 60° inner angle was selected to fabricate two FETs along two axes. HfO$_2$ was then deposited with a thickness of 15 nm as the top dielectric, followed by the fabrication of two top gate electrodes (30 nm Au). The optical image of a typical inverter device is shown in Fig. 4b. The transfer curves along the two directions are shown in the inset of Fig. 4c, which confirms the anisotropic behavior and determines two axes as well. The circuit diagram is shown in the inset of Fig. 4b, where the top gate voltage on the a-axis is fixed at -2 V, the top gate voltage on the b-axis is the input voltage $V_{in}$, and the middle shared electrode is the output voltage $V_{out}$. Fig. 4c shows the voltage transfer characteristics with excellent logic-level conservation of our digital inverter while $V_{in}$ varies between -4 V and 2 V for three different $V_{DD}$ values (3 V, 2 V, and 1 V with different colors). When $V_{in}$ is above -1 V, $V_{out}$ approaches 0 V, which denotes a digital 0. Similarly, when $V_{in}$ is below -3 V, $V_{out}$ approaches $V_{DD}$, which denotes a digital 1. The output swing (defined as the largest difference of $V_{out}$) is close to the supply voltage $V_{DD}$. As the most important parameter of a digital inverter, the gain is defined as $|dV_{out}/dV_{in}|$, which represents the sensitivity of $V_{out}$ to the change in $V_{in}$. For this device, the gain reaches 4.4 when $V_{DD}$=3 V (as shown in Fig. 4d). The gain that we obtained is clearly larger than the unity gain (gain = 1), which is required in integrated circuits consisting of multiple cascaded inverters, such as ring oscillators, and is comparable to the MoS$_2$ based inverters[29,30].

In conclusion, we have systematically studied the anisotropic properties of atomically thin ReS$_2$, a semiconducting 2D TMD with a distorted 1T structure. We fabricated mono- and few-layer ReS$_2$ FET devices that exhibit competitive performance, including a high current on/off ratio (~10$^7$) and a steep subthreshold swing (100 mV dec$^{-1}$) at room temperature. We also obtained an anisotropic ratio along the two principle axes of ReS$_2$ that is the highest obtained for all experimentally studied 2D materials. Finally, we successfully demonstrated a digital inverter device by integrating two anisotropic ReS$_2$ FET devices. Our results underscore the unique properties of semiconducting 2D materials with low symmetry, which can be exploited for novel applications in future electronics.

During the preparation of this manuscript, we became aware of another work studying the FET properties of few-layer ReS$_2$[33].

## Methods

### Materials and devices

Single crystals of ReS$_2$ were grown by the same Br$_2$-assisted chemical vapor transport method described in Ref. [17]. We used a standard mechanical exfoliation method to isolate mono- and few-layer ReS$_2$ films. The number of layers can be identified by using the color interference of a 285 nm-thick SiO$_2$ wafer and further confirmed by measuring the thickness of the flakes by using a Bruker Multimode 8 atomic force microscope (AFM). Micro Raman scattering experiments (Horiba-JY T64000) were carried out under ambient conditions in the backscattering geometry. The incident laser wavelength was 514.5 nm, and the power was less than 1 mW to minimize laser heating. Due to the limitations of the spectrometer, we only measured Raman modes above 100 cm$^{-1}$.

A conventional electron-beam lithography process (FEI F50 with Raith pattern generation system) followed by standard electron-beam evaporation of metal electrodes (typically 5 nm Ti/ 50 nm Au) was used to fabricate mono- and few-layer ReS$_2$ FET devices.

### *Ab initio* calculations

Band structures for mono- and few-layer ReS$_2$ were calculated using the generalized gradient approximation (GGA)-Perdew-Burke-Ernzerhof (PBE) function, as implemented in the VASP code (Vienna ab-initio Simulation Package) within the DFT[34,35]. Projector-augmented wave potentials were adopted[36]. The kinetic energy cutoff for the plane-wave basis set was 550 eV. The few-layer ReS$_2$ was simulated with a vacuum layer of 15 Å in the interlayer direction to ensure negligible interaction between its periodic images. In the self-consistent calculations, the Brillouin zone integration was performed on uniform Monkhorst-Pack of 24 × 24 × 1 for few-layer

ReS$_2$ and 24 × 24 × 24 for bulk ReS$_2$. The convergence criterion of self-consistent calculations for ionic relaxations was 10$^{-5}$ eV between two consecutive steps. The internal coordinates and lattice constants were optimized until the atomic forces became less than 0.001 eVÅ$^{-1}$ and the pressures on the lattice unit cell became less than 0.5 kbr.

## Acknowledgements

We thank Chun Ning Lau, Marc Bockrath and J. Joshua Yang for stimulating discussions. This work was supported in part by the National Key Basic Research Program of China (2015CB921600, 2013CBA01603), the National Natural Science Foundation of China (11374142), the Natural Science Foundation of Jiangsu Province (BK20130544, BK20140017), the Specialized Research Fund for the Doctoral Program of Higher Education (20130091120040), and Fundamental Research Funds for the Central Universities. H. T. Y., H. Y. H., and Y. C. were supported by the Department of Energy, Office of Basic Energy Sciences, Division of Materials Sciences and Engineering, under contract DE-AC02-76SF00515.

## Author contributions

E. L. and Y. F. contributed equally to this work. F. M. and H. T. Y. conceived the project and designed the experiments. E. L., Y. F. and Y. W. carried out the device fabrication and electrical measurements. Y. F., H. L. and X. W. carried out the DFT calculations. W. Z. performed the Raman spectroscopy measurements and analysis. Y. S. H. and C. H. H. performed the $ReS_2$ single crystal growth. Z. C., L. W. and A. L. performed the $HfO_2$ growth. F. M., H. Y., B. W., E. L. and Y. F. performed the data analysis and interpretation. F. M., E. L., Y. F. and H. T. Y. co-wrote the paper, with all authors contributing to the discussion and preparation of the manuscript.

## Competing financial interests

The authors declare that they have no competing financial interests.

**Figure Captions**

**Figure 1. Characterization and band structure of thin-layer ReS$_2$. a**, Crystal structure of mono-layer ReS$_2$ with a side view in the top panel and a top view in the bottom panel. Both directions of a- and b-axes are denoted by red arrows. **b**, Optical image of a mono-layer ReS$_2$ flake. The scale bar is 10 μm. **c**, AFM image of a mono-layer ReS$_2$ flake. The scale bar is 1 μm. Inset: height profile along the blue line indicating a single layer. **d**, Micro Raman experimental results performed on mono-layer, five-layer and bulk ReS$_2$. Six labeled Raman modes include two low frequency A$_g$-like modes corresponding to the out-of-plane vibrations of Re atoms and four E$_g$-like modes corresponding to the in-plane vibrations of Re atoms. The rest 12 higher frequency Raman modes are vibrations mainly from lighter S atoms. **e**, Band structure of mono-, tri- and five-layer ReS$_2$ by *ab initio* calculations indicating band gaps of 1.44, 1.40 and 1.35 eV, respectively.

**Figure 2. ReS$_2$ field effect transistor devices. a**, Transfer curves of mono- (red) and tri-layer (blue) ReS$_2$ FET devices. $V_{ds}$ is fixed to 100 mV. The on/off ratio is approximately $10^7$ for the mono-layer device and $10^7$ for the seven-layer device. The subthreshold swings are 310 mV dec$^{-1}$ (mono-layer) and 100 mV dec$^{-1}$ (tri-layer), respectively. Inset: optical image of a typical mono-layer ReS$_2$ FET device. The scale bar is 5 μm. **b**, $I_{ds}$-$V_{ds}$ curves of a mono-layer ReS$_2$ FET at different $V_{bg}$, with linear dependence indicating the ohmic contact. **c**, The dependence of device mobility on the number of layers. In general, the mobility increases monotonically with the number of layers with some scattering.

**Figure 3. Anisotropic properties of ReS$_2$. a**, Optical image of a typical thin ReS$_2$ flake with a quadrilateral shape. The scale bar is 5 μm. **b**, The statistics of inner angles for over 20 thin ReS$_2$ flakes, showing the greatest prevalence for 60° and 120°. **c**, Transfer curves of anisotropic ReS$_2$ FETs along two sides (A and B direction) of a quadrilateral-shaped five-layer flake (with an inner angle of 60º or 120º). Top inset: optical image of the devices. The scale bar is 10 μm. Low inset: the 4-probe resistance of the same devices with $V_{bg}$ varying between 0 V and 60 V. **d**, Normalized field effect mobility of a six-layer device along 12 directions evenly spaced at 30º apart plotted in polar coordinates (red dots with left axis). The direction with the lowest mobility was set to be the 0º (or 180º) reference. The optical image of the device is shown in the inset. The calculated mobility of mono-layer ReS$_2$ along three

orientations (a-axis, b-axis and perpendicular to the a-axis) is plotted in the same graph (blue dots with right axis) for comparison. The lowest mobility (a-axis) direction was set to be the 0º (or 180º) reference as well.

**Figure 4. Integrated digital inverters. a**, A schematic showing the structure of an inverter combining two top-gated anisotropic $ReS_2$ FETs. The left FET is along the a-axis, and the right FET is along the b-axis, where a Re atomic chain is highlighted in red. **b**, Optical image of a typical inverter device. A quadrilateral-shaped few-layer $ReS_2$ flake with a 60° inner angle was used to fabricate FETs along two axes covered by 15 nm-thick $HfO_2$ as the top dielectric and two top-gate electrodes (30 nm Au). The scale bar is 10 μm. Inset: the circuit diagram of the inverter, where the top-gate voltage along the a-axis is fixed at -2 V, the top gate voltage along the b-axis is the input voltage $V_{in}$, and the middle shared electrode is the output voltage $V_{out}$. **c**, Transfer characteristics of an inverter operated at $V_{DD}$ = 1 V, 2 V and 3 V. Inset: The transfer curves of two FETs with 100 mV $V_{ds}$, confirming the anisotropic behavior. **d**, The signal gain of the inverter extracted from Fig. 4c.

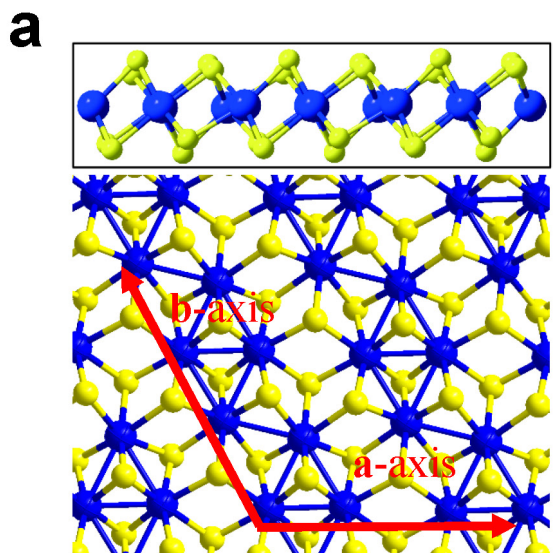
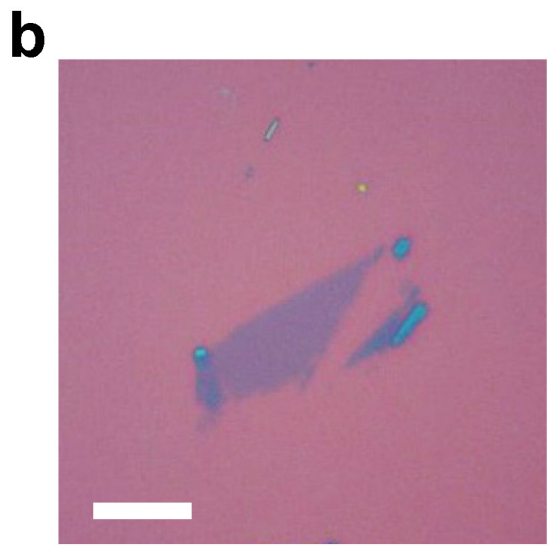
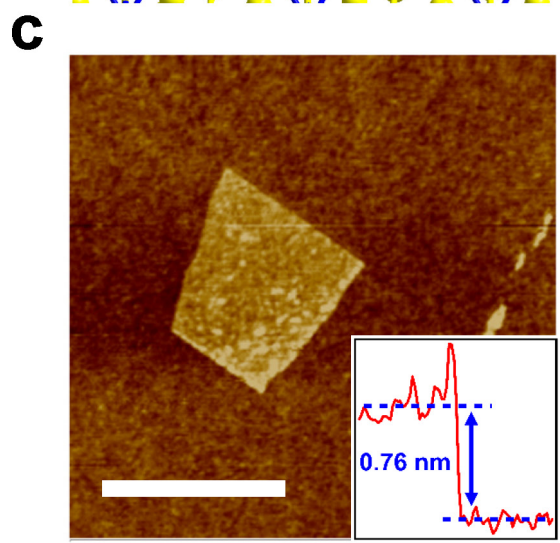
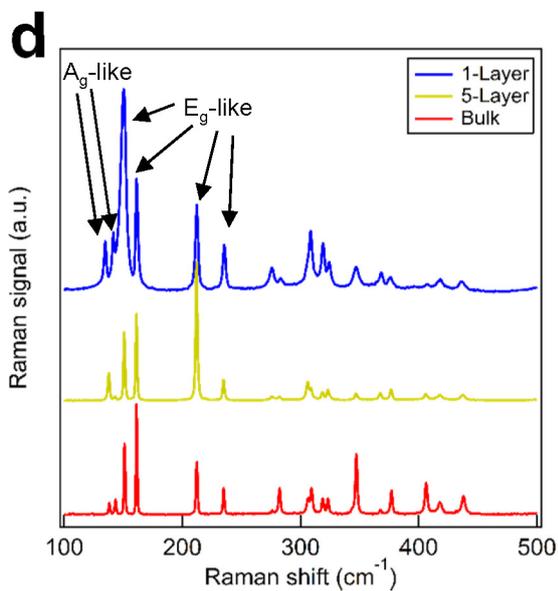
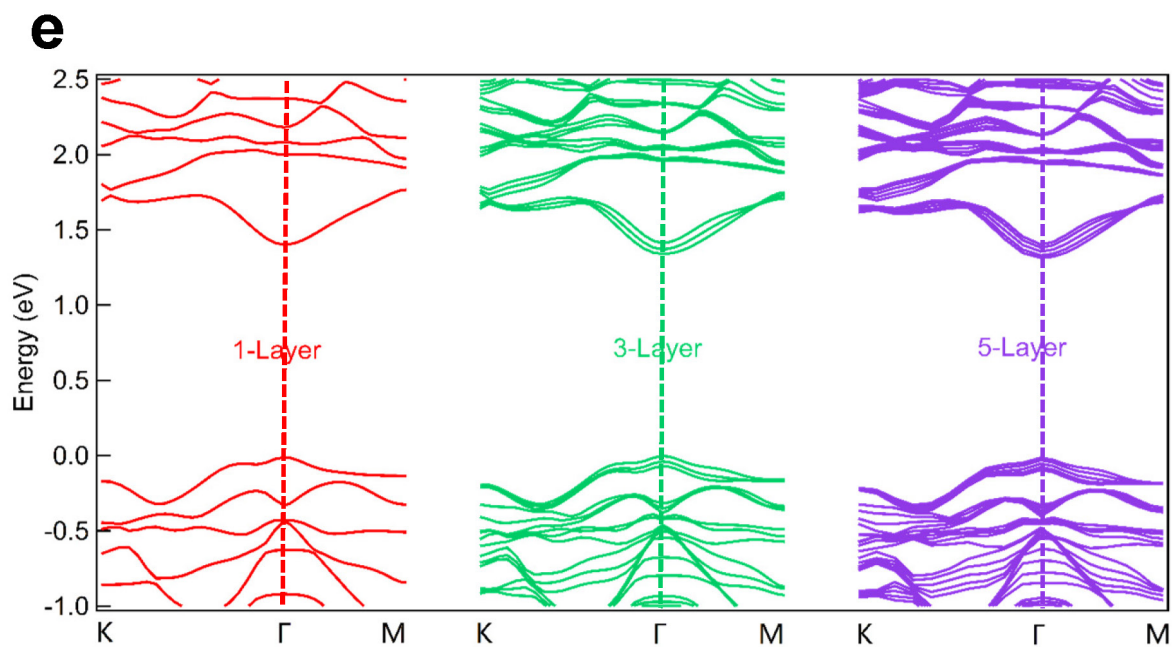

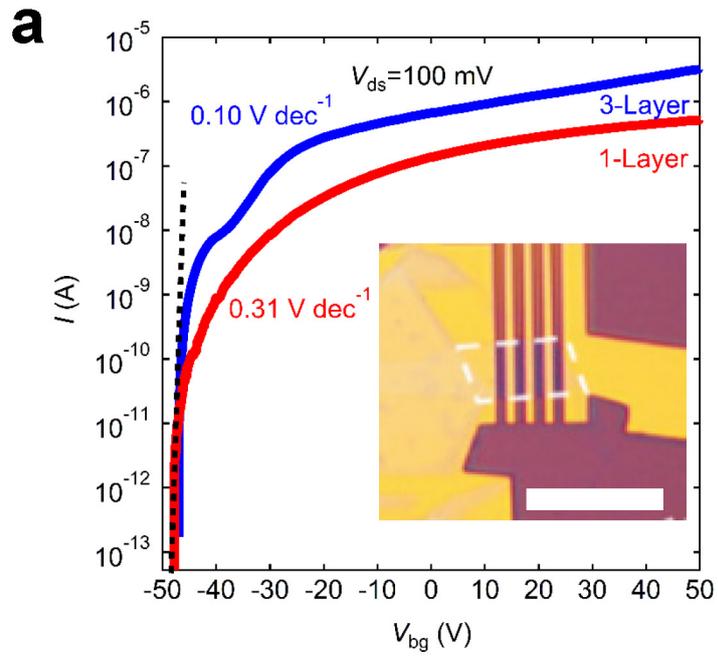

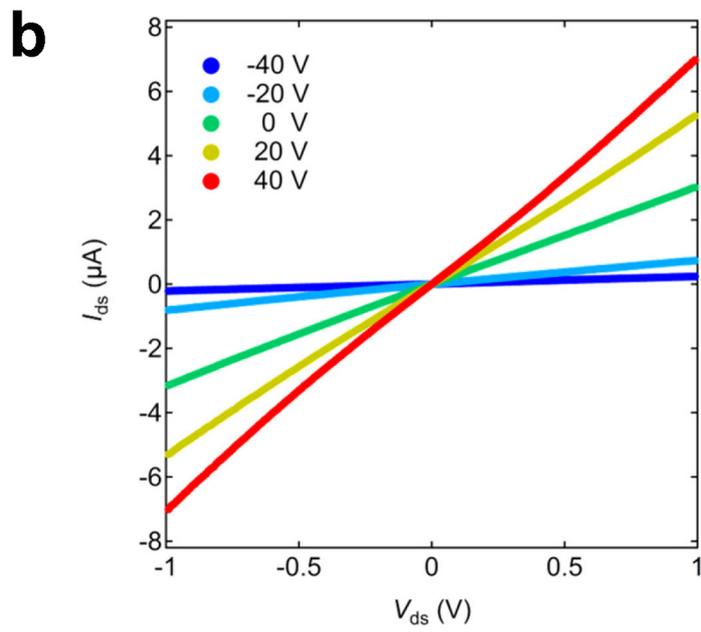

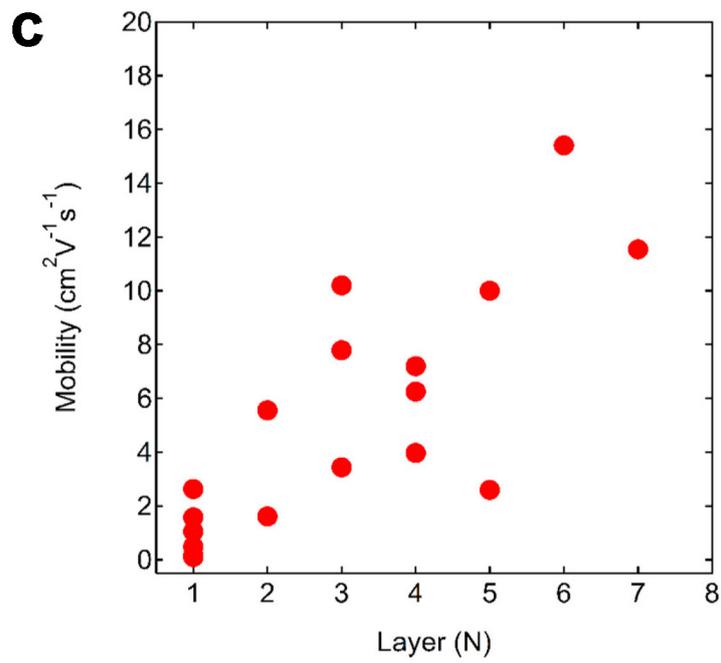

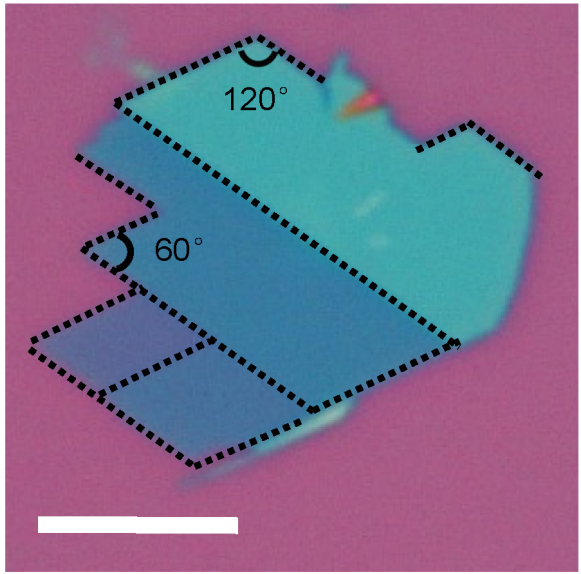
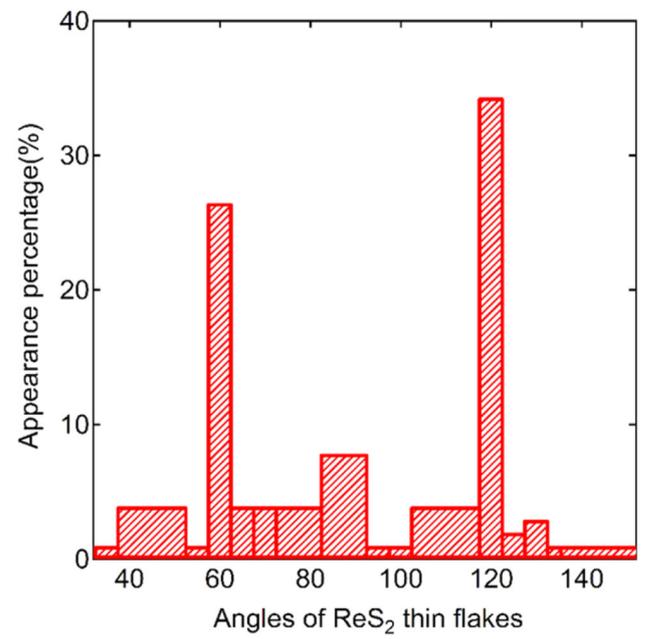
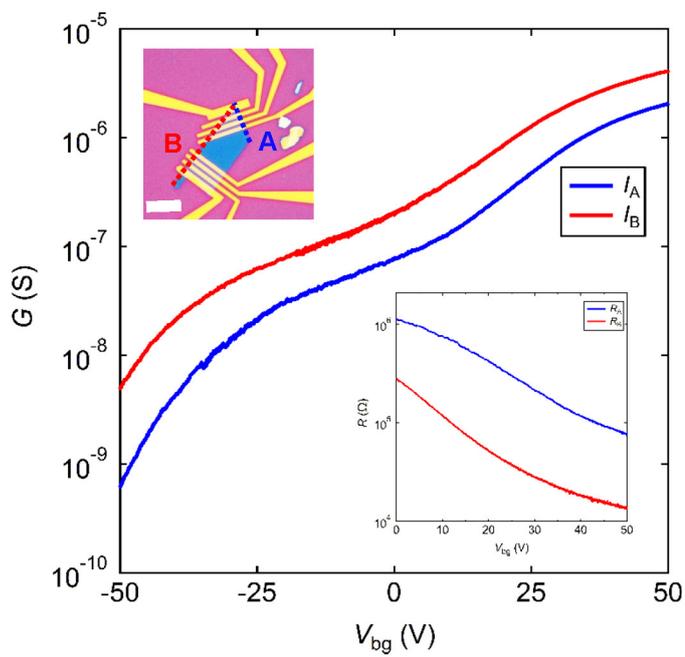
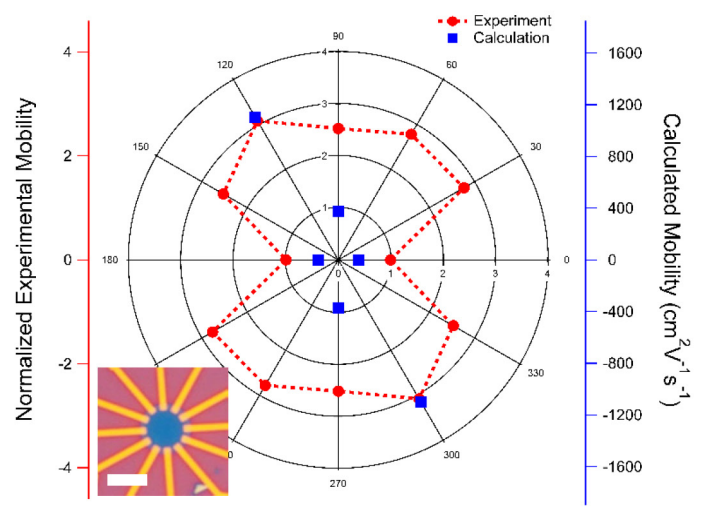

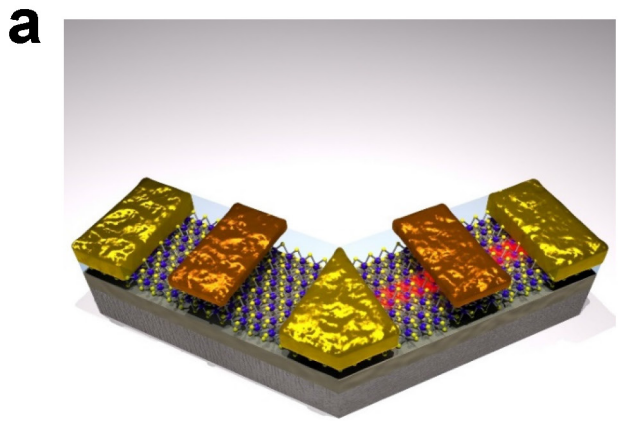
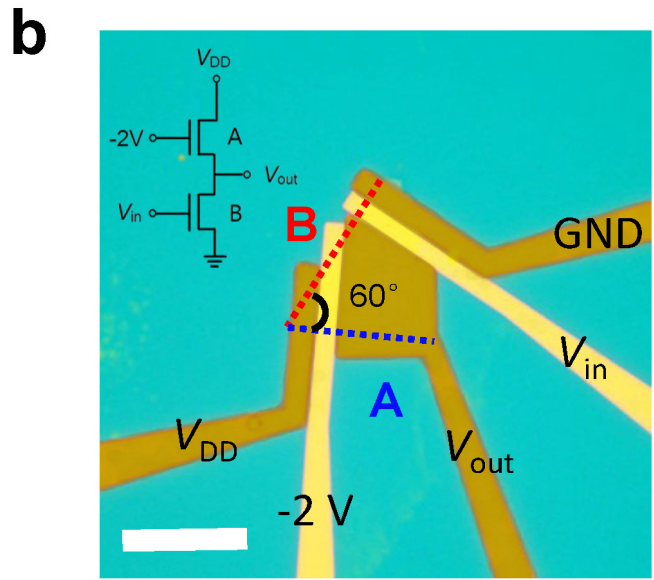
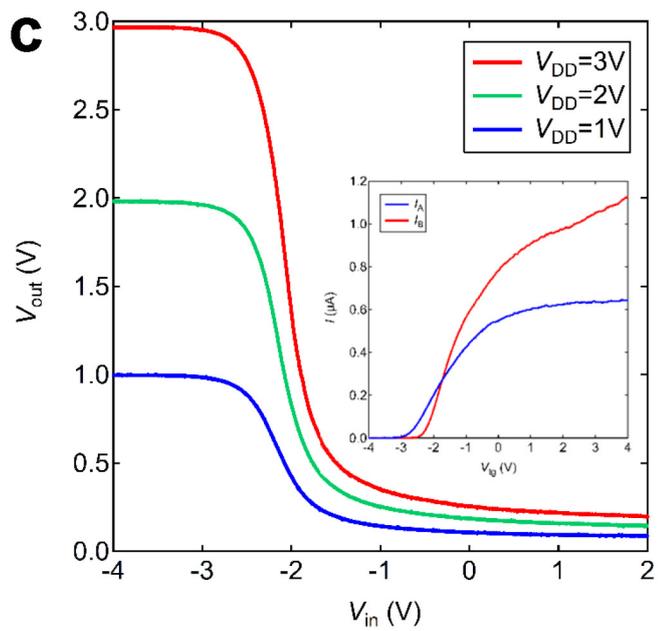
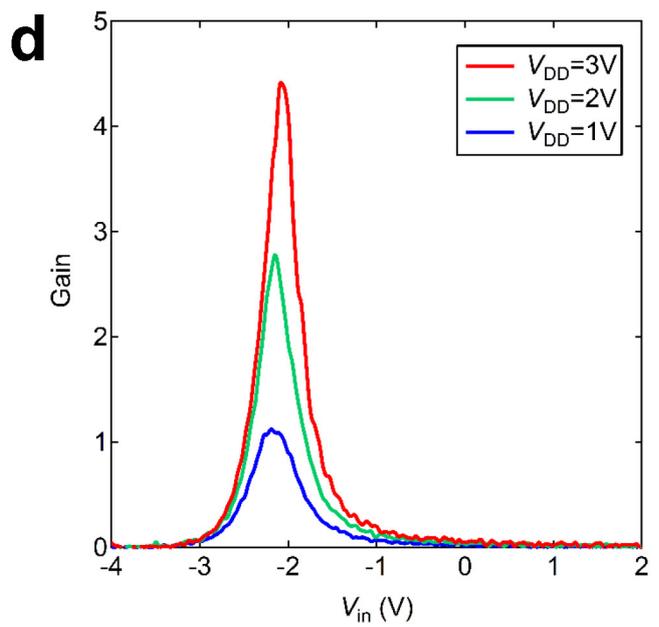

# Supplementary Figures

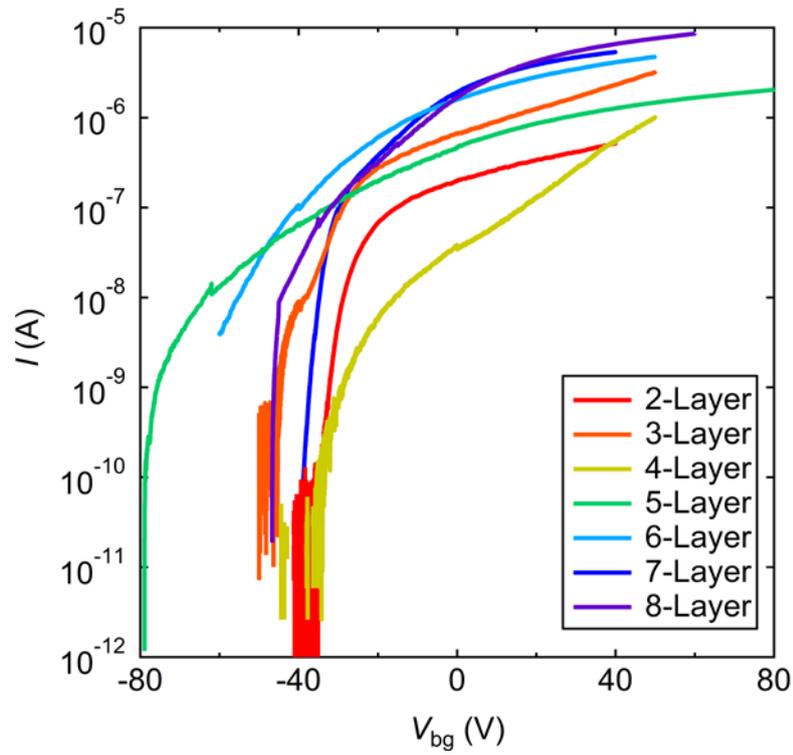

**Supplementary Figure 1: Transfer curves of few-layer ReS$_2$ FETs.** The bias voltage $V_{ds}$ was fixed at 100 mV.

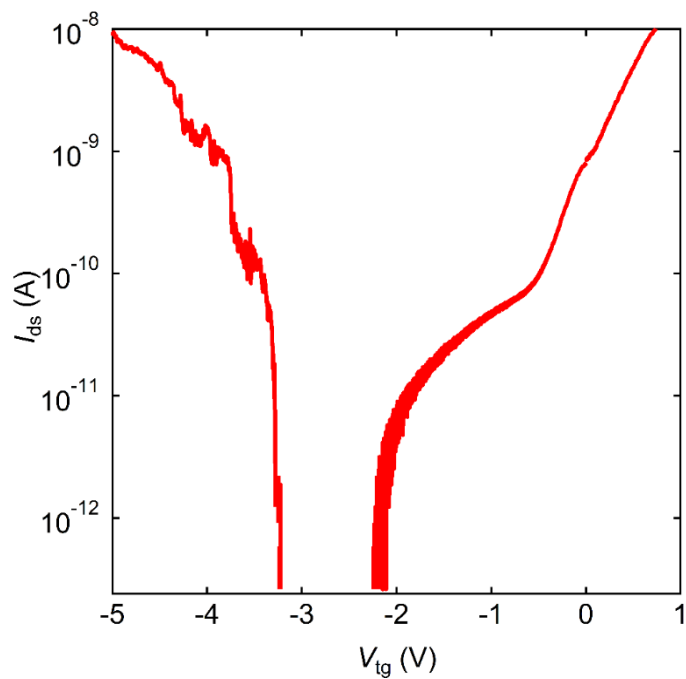

**Supplementary Figure 2: Transfer curve of an ambipolar ReS$_2$ EDLT.** The source-drain voltage was fixed at 100 mV, and the back gate remained floated. The measurements were taken at 220 K.

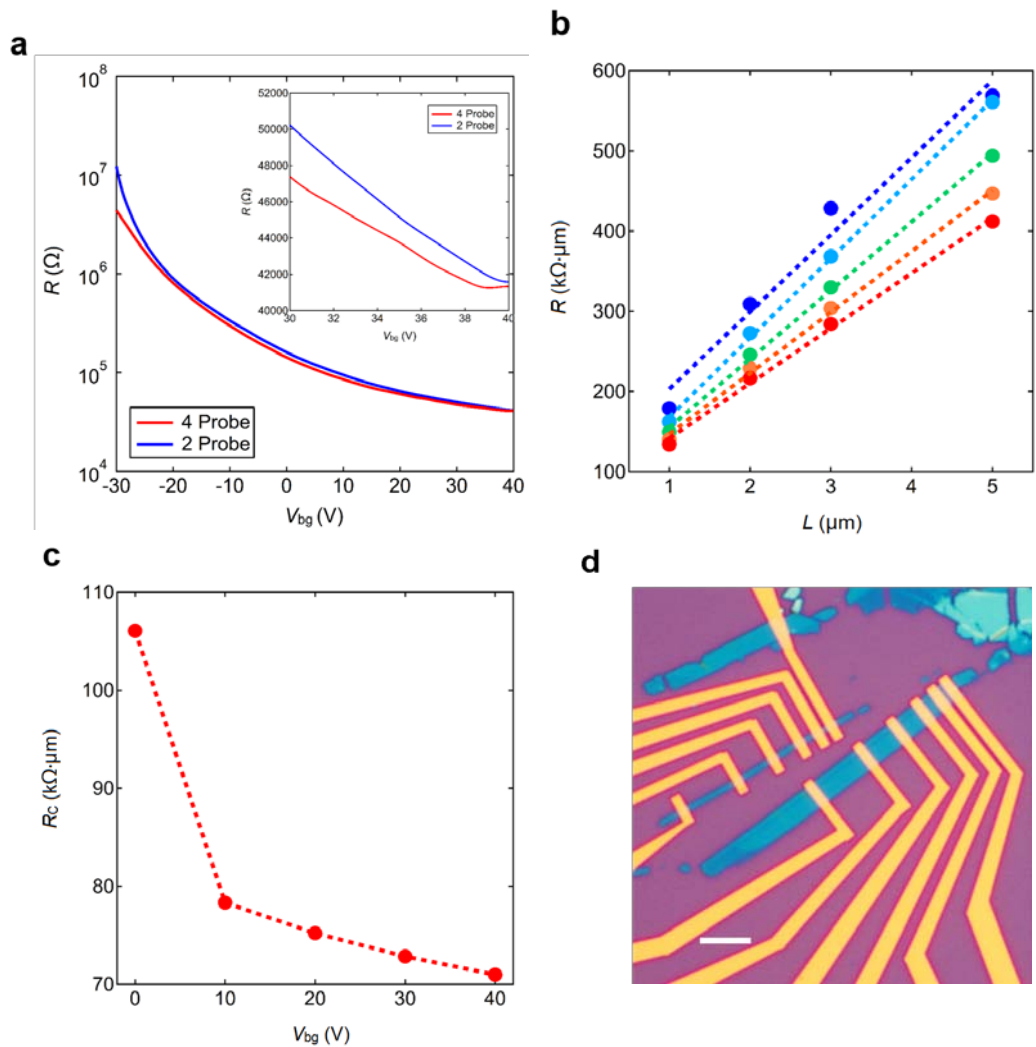

**Supplementary Figure 3: Contact of ReS$_2$ FETs.** a) Comparison of the two-probe and four-probe measurements. The inset is the zoom-in data when $V_{bg}$ varies from 30 V to 40 V. b) The channel resistance with different length under various back gate voltage. c) The contact resistance as a function of back gate voltage. d) The microscope image of the device. The scale bar is 5 μm.

## Supplementary Notes

**Supplementary Note 1: Few-layer ReS$_2$ FETs**

We measured over 40 ReS$_2$ FETs with thicknesses ranging from 0.8 to 7.5 nm (1-10 layers with an interlayer spacing of approximately 0.7 nm). All of the FETs behaved similarly as excellent n-type FET devices with the back gate swept between -80 V and +80 V. Some of the data from few-layer devices are summarized and shown in Supplementary Figure 1. The current on/off ratio can reach up to $10^7$-$10^8$, which is comparable to that of MoS$_2$ devices [1]. All measurements were carried out in a vacuum (approximately $10^{-5}$ mbar) at room temperature.

**Supplementary Note 2: Ambipolar behavior of ReS$_2$ electric double-layer transistor**

To further explore the field effect of mono- and few-layer ReS$_2$ devices, an electric double-layer transistor (EDLT) using ionic liquid gating is introduced. In our experiments, the EDLT was fabricated by dropping a droplet of an ionic liquid, N,N-diethyl-N-(2-methoxyethyl)-N-methylammonium bis(trifluoromethylsulfonyl)imide (DEME-TFSI), onto the surface of a ReS$_2$ FET device. For more efficient tuning of carrier density, the side electrode pad of the ionic liquid was designed to have a larger area than the ReS$_2$ channel of the device. To avoid possible chemical reactions between the ionic liquid and ReS$_2$, all measurements were performed at 220K and in a vacuum environment of approximately $10^{-5}$ mbar. An ambipolar behavior was observed when we swept the ionic liquid gate voltage ($V_{LG}$), with a typical transfer curve measured from a seven-layer ReS$_2$ device shown in Supplementary Figure 2. When $V_{LG}$ is above -2 V, the device behaves as an n-type transistor. When $V_{LG}$ is below -3.2 V, a p-type transistor behavior appears, indicating the shift of Fermi-level $E_F$ to access the valence band.

**Supplementary Note 3: Contact of ReS$_2$ FETs**

During the fabrication of mono- and few-layer ReS$_2$ FETs, we used 5 nm Ti covered by 50 nm Au as electrodes to make the contact. These devices have shown good contact behavior. To estimate the influence of contact resistance, we compared the measurement results obtained from both two-probe and four-probe methods. As shown in Supplementary Figure 3a, the resistance measured by the two-probe method ($R_{2P}$) is slightly larger than that of the four-probe method ($R_{4P}$). The resistance difference ($\Delta R$) of $R_{2P}$ and $R_{4P}$ should be mainly attributed to the contact resistance. If we define $\alpha = \frac{\Delta R}{R_{2P}}$ as the weight of $\Delta R$ in $R_{2P}$, we found $\alpha$ is highly gate-tunable. In the n-doped regime as our work focused on, $\alpha$ is less than 10% (when $V_{bg}$ = -20 V) and decreases notably as the doping level increases, to be as low as 1% when $V_{bg}$ = 40 V.

We also used the transfer length method to estimate the contact resistance of the ReS$_2$ FETs. As shown in Supplementary Figure 3b, we measured the channel resistance with different length under various back gate voltage $V_{bg}$. The extrapolated contact resistance $R_C$ versus $V_{bg}$ is shown in Supplementary Figure 3c. The contact resistance decreases with the increasing $V_{bg}$, ranging from 71 kΩ to 106 kΩ, which is similar to the results reported [2]. Supplementary Figure 3d shows the optical image of the multi-terminal device.

**Supplementary Note 4: Carrier mobility calculation**

Our carrier mobility calculation was performed using the VASP (Vienna ab-initio Simulation Package) code [3, 4]. The results presented in the following were obtained by using the generalized gradient approximation (GGA)-Perdew-Becke-Erzenhof (PBE) function [5], a 10×5×1 mesh for the Brillouin zone sampling and a cut-off of 500 eV for the plane-wave basis set. An orthogonal supercell was created for ReS$_2$ sheets, with the atomic plane and its neighboring image separated by a 36 Å vacuum layer. All of the structures were relaxed until the Hellmann−Feynman forces became less than 0.01 eVÅ$^{-1}$.

In 2D materials, the carrier mobility is given by the following formula [6, 7]:

$$\mu_{2D} = \frac{2e_0 \hbar^3 C_q}{3k_B T E_1^2 m^{*2}} \quad (1)$$

where $m^*$ is the effective mass and $T$ is the temperature (set to be 300K in our calculation). The term $E_1$ is the deformation potential constant, which denotes the shift in the band edges along the transport directions induced by strain $\varepsilon$ (calculated using a step of 1%). It is defined as $E_1 = \frac{\partial E_M}{\partial \varepsilon}$, where $E_M$ is the energy of the conduction band minimum for electrons and the valence band maximum for holes. $C_q$ is the stretching modulus caused by $\varepsilon$, which can be calculated by $C_q = \frac{1}{S_0} \frac{\partial^2 E_{total}}{\partial \varepsilon^2}$, where $E_{total}$ is the total energy and $S_0$ is the lattice volume at equilibrium for a 2D system.

# Supplementary References